 \documentclass[nohyper,12pt,letterpaper]{JHEP3}
 \usepackage{epsfig}
 
\title{Gluonic fields of a static particle to all orders in 1/N.}

\author{Bartomeu Fiol$^1$, Blai Garolera$^1$ and Aitor Lewkowycz$^2$  \\

$^1$Departament de F{\'\i}sica Fonamental i \\Institut de Ci{\`e}ncies del Cosmos, 

Universitat de Barcelona,

Mart{\'\i}\ i Franqu{\`e}s 1, 08028 Barcelona, Catalonia, Spain \\

$^2$Perimeter Institute for Theoretical Physics,

Waterloo, Ontario N2L 2Y5, Canada \\

\email{bfiol@ub.edu, bgarolera@ffn.ub.es, alewkowycz@pitp.ca}}

\abstract{
We determine the expectation value of the gauge invariant operator $Tr [F^2+\dots ]$ for
${\cal N}=4$ SU(N) SYM, in the presence of an infinitely heavy static particle in the symmetric representation of SU(N). We carry out the computation in the context of the AdS/CFT correspondence, by considering the perturbation of the dilaton field caused by the presence of a  D3 brane dual to such an external probe. We find that the effective chromo-electric charge of the probe has exactly the same expression as the one recently found in the computation of energy loss by radiation.}

\begin{document}

\section{Introduction and conclusions}
Among the many applications of the AdS/CFT correspondence, a very broad area of research is the study of the behavior of external probes  in strongly coupled field theories and the response of the fields to such probes. The first examples of such computations were the evaluation of the static quark-antiquark potential in \cite{Rey:1998ik, Maldacena:1998im} by means of a particular string configuration reaching the boundary of $AdS$. Following those seminal works,  the key idea of realizing external heavy quarks by strings in the bulk geometry has been generalized in many directions. In particular, as we will briefly review, probes transforming under different representations of the gauge group are holographically realized by considering different types of branes in the supergravity background.

An external probe in the fundamental representation of the gauge group is dual to a string in the bulk. At least for the simplest implementations of this identification (i.e. in the absence of additional scales like finite mass or non-zero temperature), the computed ${\cal N}=4$ SYM observables reveal a common feature: while at weak 't Hooft coupling $\lambda$ we can identify this coupling as the analogous of the charge squared, at strong 't Hooft coupling there is a screening of this charge, in the sense that the results obtained demand the strong coupling identification
\begin{equation}
e^2_{fund} \sim \sqrt{\lambda}
\label{chargefund}
\end{equation}
This generic behavior stems from the fact that the Nambu-Goto action evaluated for world-sheet metrics embedded in $AdS_5$ goes like
$$
S_{NG}=-\frac{1}{2\pi \alpha'} \int d^2 \sigma \sqrt{-|g|} =-\frac{\sqrt{\lambda}}{2\pi L^2}\int d^2 \sigma \sqrt{-|g|}
$$
where $L$ is the $AdS_5$ radius which generically cancels out from this expression when specific world-sheet metrics are plugged-in.  Some examples of this are the original quark-antiquark potential \cite{Rey:1998ik, Maldacena:1998im}, the computation of the gauge invariant operators in the presence of  a particle at rest \cite{hep-th/9812007, hep-th/9906153} or following arbitrary motion \cite{Athanasiou:2010pv, arXiv:1106.4059}, and the formula for energy loss by radiation \cite{hep-th/0305196}.

Moving on to probes transforming in other representations of the gauge group, it is by now well understood that they are realized by D3 and D5 branes. Specifically, on the gravity side, the duals of particles in the symmetric or antisymmetric representations of the gauge group are given by D3 and D5 branes respectively, with world-volume fluxes that encode the rank of the representation \cite{Gomis:2006sb, Yamaguchi:2006tq, Hartnoll:2006hr}. One of the novel features of this identification - and the chief reason for our interest in the topic - is that some computed observables are functions of $k/N$, where $k$ is the rank of the symmetric/antisymmetric representation. Thus, by means of the AdS/CFT correspondence, the physics of these probes of ${\cal N}=4$ SYM can be studied at large $N$, large $\lambda$, beyond the leading term.

While the holographic prescription is in principle equally straightforward for the study of probes in the symmetric and the antisymmetric representations, when it comes to actual computations we currently face more difficulties in the symmetric case than in the antisymmetric one. One of the reasons behind this difference comes from the existence of a quite universal result for the embedding of D5 branes in terms of embeddings of fundamental strings, due to Hartnoll \cite{Hartnoll:2006ib}: given a string world-sheet that solves the Nambu-Goto action in an arbitrary five-dimensional manifold $M$ with constant negative Ricci tensor, there is a quite general construction that provides a solution for the D5-brane action in $M\times S^5$, of the form $\Sigma \times S^4$ where $\Sigma \hookrightarrow M$ is the string world-sheet and $S^4\hookrightarrow S^5$. This gives a link between the string used to describe a particle in the fundamental representation and the D5 brane used to represent a probe in the antisymmetric representation. Moreover, this link translates into a quite robust result for the evaluation of observable quantities; for the k-antisymmetric representation, typically the strong coupling $\sqrt{\lambda}$ dependence of observables for probes in  the fundamental representation gets replaced (up to numeric factors) by 
\begin{equation}
\sqrt{\lambda} \rightarrow \frac{2N}{3\pi}\sin^3 \theta_0 \sqrt{\lambda} \sim  e^2_{A_k}
\label{chargeanti}
\end{equation}
where $\theta_0$ denotes the angle of $S^4$ inside $S^5$ and is the solution of \cite{hep-th/0104082}
$$
\sin \theta_0 \cos \theta_0 -\theta_0=\pi\left(\frac{k}{N}-1\right)
$$
This identification is supported by explicit computations of Wilson loops \cite{Yamaguchi:2006tq,Hartnoll:2006ib} which match matrix model computations \cite{Hartnoll:2006is}, energy loss by radiation in vacuum \cite{fiol11} and in a thermal medium \cite{Chernicoff:2006yp}, or the impurity entropy in supersymmetric versions of the Kondo model \cite{arXiv:1012.1973}.

On the other hand, for probes in the symmetric representation we currently don't have a generic construction that links the string that realizes a particle in the fundamental representation with a D3 brane that realizes a probe in the symmetric representation. Furthermore, while the observables analyzed so far seem to depend on the combination
\begin{equation}
\kappa=\frac{k\sqrt{\lambda}}{4N}
\label{defkappa}
\end{equation}
introduced in \cite{Drukker:2005kx}, they do not display a common function that replaces the $\sqrt{\lambda}$ dependence of the fundamental representation. For instance, in the computation of the energy loss by radiation in vacuum of a particle moving with constant proper acceleration, it was found in \cite{fiol11} that
\begin{equation}
\sqrt{\lambda} \rightarrow 4N\kappa \sqrt{1+\kappa^2} = k\sqrt{\lambda}\sqrt{1+\frac{k^2 \lambda}{16 N^2}} \stackrel{?}{\sim} e^2_{S_k}
\label{chargesym}
\end{equation}
while for the vev of a circular Wilson loop, it was found that \cite{Drukker:2005kx}
$$
\sqrt{\lambda} \rightarrow 2N (\kappa \sqrt{1+\kappa^2}+ \sinh^{-1} \kappa)
$$
While both functions expanded as a power series  in $\kappa$ start with the common term $k\sqrt{\lambda}$ (i.e. $k$ times the result for the fundamental representation) they are clearly different beyond this leading order.

The purpose of this note is to shed some light on the issue of observables for probes in the symmetric representation, by computing the expectation value of a particular gauge invariant operator in the presence of an infinitely heavy half-BPS static particle, transforming in the k-symmetric representation of ${\cal N}=4$ $SU(N)$ SYM. More specifically, this operator is the one sourced by the asymptotic value of the dilaton \cite{hep-th/9702076}\footnote{We follow the conventions of \cite{arXiv:1106.4059}.}, 
$$
{\cal O}_{F^2}=\frac{1}{2 g_{YM}^2} \hbox{Tr }\left(F^2 +[X_I,X_J][X^I,X^J] \; \hbox{+ fermions }\right)
$$
On general grounds, in the presence of a static probe placed at the origin, we expect the one-point function to be of the form
$$
<{\cal O}_{F^2}(\vec x)>=\frac{f(k,\lambda, N)}{|\vec x|^4}
$$ 
and our objective is to compute the  dimensionless function $f(k,\lambda, N)$ when the probe transforms in the k-symmetric representation of $SU(N)$. By analogy with the Coulombic case, one might refer to $f(k,\lambda, N)$ as the square of the "chromo-electric charge" of the heavy particle. To carry out this computation we will consider a particular half-BPS D3-brane embedded in $AdS_5\times S^5$ and analyze the linearized perturbation equation for the dilaton, with the D3-brane acting as source. The advantage of considering this operator is that the perturbation equation of the dilaton decouples from the equations for metric perturbations, so its study is quite straightforward.

The analogous computation for a particle in the fundamental representation was carried out some time ago, considering in that case the perturbation equation for the dilaton sourced by a fundamental string \cite{hep-th/9812007, hep-th/9906153} (see also \cite{arXiv:1106.4059}). For the sake of comparison, let's quote their final result in our conventions,
$$
<{\cal O}_{F^2}(\vec x)>_{fund}\; = \; \frac{\sqrt{\lambda}}{16\pi^2}\frac{1}{|\vec x|^4}
$$
In the next section we will give details of our computation, but let us jump ahead and present the final result,
$$
<{\cal O}_{F^2}(\vec x)>_{S_k}\; =\; \frac{N\kappa \sqrt{1+\kappa^2}}{4\pi^2}\frac{1}{|\vec x|^4}=\frac{k\sqrt{\lambda}}{16\pi^2}\frac{\sqrt{1+\frac{k^2 \lambda}{16 N^2}}}{|\vec x|^4}
$$
As we can see, we obtain again a result organized as a function of $\kappa$. Furthermore, the form of the function replacing the $\sqrt{\lambda}$ result of the fundamental representation is exactly the same (including the numerical factor) that was found in the computation of the energy loss \cite{fiol11}. This supports the identification of the effective charge for probes in the symmetric representation made in eq. (\ref{chargesym}). To us, the complete coincidence of these two results was far from a foregone conclusion: while the two D3-brane solutions used in the computations are related by a conformal transformation, this doesn't even imply that the {\it same} observable will coincide, as the differing expectation values of the corresponding Wilson loops beautifully show \cite{Drukker:2000rr, Drukker:2005kx}. Moreover, the energy loss computation in \cite{fiol11} captured physics of radiative fields, encoded in the bulk by the presence of a horizon 
in the world-volume  of the D3-brane, while in the computation to be presented shortly, the physics of static fields is captured by the behavior near the $AdS$ boundary, and the 
D3-brane world-volume has now no horizon.\footnote{Incidentally, the world-volume metric of the D3-brane we will consider is $AdS_2\times S^2$ with radii $L\sqrt{1+\kappa^2}$ and $L\kappa$ respectively \cite{Drukker:2005kx}, so $\kappa \sqrt{1+\kappa^2}$ (divided by $L^2$) happens to be the product of these two radii.}

Let us conclude this introduction by mentioning some of the questions we would like to address in the future. For probes in the fundamental and the antisymmetric representations, a variety of observables at strong coupling (even without supersymmetry and at non-zero temperature!) support the identification of effective charges in eqs. (\ref{chargefund}) and (\ref{chargeanti}) respectively, for probes following arbitrary world-lines. For the symmetric representation, a first limitation is that currently we only have at our disposal the D3 brane corresponding to a static probe in the vacuum and the conformally related case of hyperbolic motion. It would be highly desirable to find D3 branes that correspond to probes following arbitrary world-lines.

Even for this very limited set of world-lines, there are other gauge invariant operators whose expectation value would be interesting to compute, {\it e.g.} the energy-momentum tensor of the gluonic fields in the presence of the static probe considered here, along the lines of \cite{Athanasiou:2010pv}, or perhaps more interestingly, for the same probe in hyperbolic motion \cite{fiol11, Drukker:2005kx}. We have shown in this note that, while there is no unique function of $\kappa$ appearing in different observables, at least for the Li\'enard-type formula for energy-loss and the expectation value of $<{\cal O}_{F^2}>$  of Coulomb-type fields, the functions of $\kappa$ coincide. In general, are there relations we can expect or prove for the $\kappa$ dependence of different observables? Finally, another interesting issue is the actual range of validity of these different computations. As argued in \cite{Drukker:2005kx, fiol11}, the {\it a priori} range of validity of the computations of the vev of the circular Wilson loop and of energy loss by radiation do not include taking $k=1$ (i.e. the fundamental representation). Nevertheless, for the case of the vev of the circular Wilson loop it was shown in \cite{Drukker:2005kx} that if one sets $k=1$ in the probe D3-brane result, one still correctly recovers (at leading order in $1/\sqrt{\lambda}$ and to all orders in $1/N$) the gauge theory result, available thanks to a matrix model computation \cite{Drukker:2000rr}. This is presumably due to the large amount of supersymmetry of the configuration, also present in the computation of energy loss in hyperbolic motion and in the current one for a static 
half-BPS particle, so it would be important to understand for which of these computations there are non-renormalization theorems that extend their range of validity beyond the region argued in \cite{Drukker:2005kx}. 

\section{The computation}
In this section we will present the details of the computation of $<{\cal O}_{F^2}>$, the expectation value of the operator that the dilaton couples to, in the presence of a heavy probe transforming in the symmetric representation of SU(N). We will first compute the linearized perturbation of the dilaton field caused by the D-brane probe, and from its behavior near the boundary of $AdS_5$ we will then read off the expectation value of ${\cal O}_{F^2}$. Our computations will closely follow the ones presented in \cite{hep-th/9812007, hep-th/9906153} (see also \cite{arXiv:1106.4059}) for the case of a probe in the fundamental representation.

We work in Poincar\'e coordinates and take advantage of the spherical symmetry of the problem
$$
ds^2_{AdS_5}=\frac{L^2}{z^2}\left(dz^2-dt^2+dr^2+r^2 d\theta^2+r^2\sin^2 \theta d\varphi^2\right)
$$
The D3-brane we will be interested in was discussed in \cite{Rey:1998ik,Drukker:2005kx}. It reaches the boundary of AdS ($z=0$ in our coordinates) at a straight line $r=0$, which is the world-line of the static dual particle placed at the origin. Since we let the D3 brane reach the boundary, the static particle is infinitely heavy. To describe the D3-brane, identify $(t,z,\theta,\varphi)$ as the world-volume coordinates; then the solution is given by a function $r(z)$ and a world-volume electric field
$$
r=\kappa z\hspace{1cm}F_{tz}=\frac{\sqrt{\lambda}}{2\pi}\frac{1}{z^2}
$$
with $\kappa$ as defined in eq. (\ref{defkappa}). As shown in detail in \cite{Drukker:2005kx} this D3-brane is half-BPS. 

Our next step is to consider at linear level the backreaction that this D3-brane induces on the $AdS_5\times S^5$ solution of IIB supergravity. More specifically, since the dilaton is constant in the unperturbed solution, and its stress-energy tensor is quadratic, at the linearized level the equation for the perturbation of the dilaton decouples from the rest of linearized supergravity equations. As in \cite{hep-th/9812007, hep-th/9906153}, we work in Einstein frame, and take as starting point the action
$$
S=-\frac{\Omega_5L^5}{2\kappa_{10}^2}\int d^5x \sqrt{-|g_E|}\frac{1}{2}g_E^{mn}\partial_m \phi \partial_n \phi
-T_{D3}\int d^4\xi \sqrt{-|G_E+e^{-\phi/2}2\pi \alpha' F|}
$$
The resulting equation of motion can be written
$$
\partial_m\left(\sqrt{-|g_E|} g_E^{mn}\partial _n \phi\right)=J(x)
$$
with the source defined by the D3-brane solution
$$
J(x)=\frac{T_{D3} \kappa_{10}^2}{\Omega_5 L} \frac{\kappa \sin \theta}{z^2} \delta\left(r-z\kappa\right)
$$
To compute $\phi(x)$ we will use its Green function $D(x,x')$
$$
\phi(x)=\int d^5x'\; D(x,x')\; J(x')
$$
It is convenient to write $D(x,x')$ purely in terms of the invariant distance $v$ defined by
\begin{equation}
\cos v=1-\frac{(t-t')^2-(\vec x-\vec x')^2-(z-z')^2}{2zz'}
\label{invav}
\end{equation}
The explicit expression for $D(v)$ can be found for instance in \cite{hep-th/9812007}
$$
D=\frac{-1}{4\pi^2L^3 \sin v}\frac{d}{dv}\left(\frac{\cos 2v}{\sin v} \Theta(1-|\cos v|)\right)
$$
To carry out the integration, we follow the same steps as \cite{hep-th/9906153}. We first define a rescaled dilaton field,
$$
\tilde \phi\equiv \frac{\Omega_5 L^8}{2\kappa_{10}^2}\phi
$$
and use eq. (\ref{invav}) to change variables from $t'$ to $v$ to obtain, after an integration by parts
$$
\tilde \phi=\frac{N\kappa z^2}{16 \pi^4}\int _0^\infty dr'  \int_0^\pi d\theta' \sin \theta'\int_0^{2\pi} d\varphi' \int _0^\infty \frac{dz' \; \delta(r'-z'\kappa)}{\left(z^2+z'^2+(\vec x-\vec x')^2\right)^{\frac{3}{2}}}\int _0 ^\pi \frac{dv\; \cos 2v}{\left(1-\frac{2zz'\; \cos v}{z^2+z'^2+(\vec x-\vec x')^2}\right)^{\frac{3}{2}}} 
$$
The integral over $v$ is the same one that appeared in the computation of the perturbation caused by a string dual to a static probe \cite{hep-th/9906153}. The novel ingredient in the computation comes from the non-trivial angular dependence in the current case. While it might be possible to completely carry out this integral, at this point it is pertinent to recall that to compute the expectation value of the dual field theory operator, we only need the leading behavior of the perturbation of the dilaton field near the boundary of $AdS_5$. Specifically \cite{hep-th/9812007}, 
\begin{equation}
<{\cal O}_{F^2}>=-\frac{1}{z^3}\partial_z \tilde \phi |_{z=0}
\label{onepoint}
\end{equation}
so for our purposes it is enough to expand the integrands in powers of $z$, and keep only the leading $z^4$ term. This simplifies the task enormously, and reduces it to computing some straightforward integrals. Skipping some unilluminating steps we arrive at
$$
\tilde \phi= \frac{N\kappa}{16 \pi^2}\frac{z^4}{(z^2+r^2)^2}\frac{1}{(1+\kappa^2)^{3/2}}\frac{1}{\left(1-\frac{\kappa^2}{1+\kappa^2}\frac{r^2}{r^2+z^2}\right)^2}+O(z^5)
$$
which upon differentiation, and setting then $z=0$ as required by eq. (\ref{onepoint}), leads to our final result
$$
<{\cal O}_{F^2}>=\frac{N\kappa \sqrt{1+\kappa^2}}{4\pi^2}\frac{1}{|\vec x|^4}=\frac{k\sqrt{\lambda}}{16\pi^2}\frac{\sqrt{1+\frac{k^2 \lambda}{16 N^2}}}{|\vec x|^4}
$$

\section{Acknowledgements} 
We would like to thank Mariano Chernicoff for helpful conversations. The research of BF is supported by a Ram\'{o}n y Cajal fellowship, and also by MEC FPA2009-20807-C02-02, CPAN CSD2007-00042, within the Consolider-Ingenio2010 program, and AGAUR 2009SGR00168. The research of BG is supported by an ICC scholarship and by MEC FPA2009-20807-C02-02. The research of AL was partly supported by a Spanish MEPSYD fellowship for undergraduate students. AL further acknowledges support from Fundaci\'on Caja Madrid and the Perimeter Scholars International program.

\end{document}